\documentclass[12pt]{article}
\input epsf.sty

\def\lromn#1{\uppercase\expandafter{\romannumeral#1}}

\usepackage{graphicx,amsmath,amssymb}
\begin{document}

\begin{center}
\begin{large}
\renewcommand{\thefootnote}{\fnsymbol{footnote}}
\textbf{
Parity violating observables
in radiative neutrino pair emission
from metastable atoms
}\footnote[4]
{Work supported in part by the Grant-in-Aid for Science Research from
the Ministry of 
Education, Science and Culture of 
Japan No. 19204028, No.20740138 and No.21244032}

\end{large}
\end{center}

\vspace{2cm}
\begin{center}
\begin{large}
M. Yoshimura, A.Fukumi, N. Sasao, and T. Yamaguchi

Center of Quantum Universe, Okayama University, \\
Tsushima-naka 3-1-1, Okayama,
700-8530 Japan

\end{large}
\end{center}

\vspace{4cm}
\begin{center}
\begin{Large}
{\bf ABSTRACT}
\end{Large}
\end{center}

We report on a possibility of measuring 
parity violating effects in radiative neutrino pair emission
from metastable atoms;
asymmetric angular distribution of emitted
photons from oriented atoms and
emergent circular polarization.
Their observation,
along with the continuous photon energy spectrum which has
6 thresholds, 
may be interpreted as events being
a combined weak and QED process,
emission of $\gamma \nu_i \nu_j$ in the final state.
The method may greatly help to perform neutrino
mass spectroscopy using atoms,
a systematic determination of the neutrino mass matrix.

\newpage
{\bf Introduction}
\hspace{0.5cm}
Despite exciting experimental developments
in neutrino physics
the neutrino is still a very mysterious particle.
Do neutrinos obey the same type of Dirac equation as all
other charged leptons and quarks, or
do they obey two-component Majorana equation in
which particle is indistinguishable from its antiparticle ?
How small is the lightest neutrino mass ?
These questions are linked to the
cosmological scenario of leptogenesis \cite{leptogenesis}
which ultimately aims at a deep understanding 
of the matter-antimatter imbalance in our universe.

Radiative neutrino pair emission (RNPE)
\cite{my-06}, \cite{rnpe-macro}
$|i \rangle_A \rightarrow | f\rangle_A + \gamma
+\nu_i\nu_j$ ($\nu_i$ being the
i-th mass eigenstate ordered in mass
by $0 < m_1 < m_2 < m_3$) from a metastable atomic state $|i \rangle_A$,
a process yet to be discovered,
gives the opportunity of systematically measuring
the neutrino mass matrix (described by 9 real parameters)
and the mass type, Majorana vs Dirac.
In the present work we calculate the magnitude of observables that evidently
indicates parity violation, providing, along with
a missing energy suggestive of invisible neutrino emission, 
a fundamental tool of unambiguous  detection of the
weak RNPE process.
The RNPE process furthermore gives critical information 
on three neutrino masses
and the unknown mixing angle $\theta_{13}$.
This way one may advance a step towards
verifying important ingredients of leptogenesis
theory.

The first experimental discovery of
parity violation was made by measuring emitted electron
correlation with polarization axis of target in the $\beta$ 
and $\mu$ decay \cite{commins}.
The measured electron in the $\beta$ 
and $\mu$ decay is replaced here in RNPE
by the photon.
In RNPE the circular polarization of emitted photon
is another good quantity for verification of
parity violation.
Mass thresholds of invisible neutrino pair emission of mass
eigenstates $\nu_i\nu_j$
are measurable in atomic experiments
by a spectrometer with good energy resolution,
while in $\beta$ and $\mu$ decays only flavor eigenstates 
$\nu_e, \nu_{\mu}$ are observable with poor energy resolution.

In order to increase feasibility of actual experiments
one needs enhancement of RNPE rates, for which we use
macro-coherence of target atoms \cite{rnpe-macro}.
Measurement of parity violation helps much to
reduce QED backgrounds even for smaller rates, which 
thus becomes important together with other means of the background rejection
\cite{ys-09}.

\vspace{0.5cm}
{\bf  Prepared states}
\hspace{0.5cm}
In discussions that follow it is essential
to have the momentum conservation 
$\vec{k} + \vec{p}_1 + \vec{p}_2= 0$ among
3 emitted light particles, $\vec{k}$ the momentum of
photon, $\vec{p}_i$ those of neutrinos, taking
infinitely heavy atom, along with
the usual energy conservation.
This way the macro-coherent decay is much similar,
in kinematics, to the ordinary decay of a single relativistic particle
of mass $\Delta$ where $\Delta = E_i - E_f$ 
is the energy difference 
of initial and final atomic states.
For the momentum-conserving three-body decay of RNPE the energy spectrum
of photon has 6 different thresholds which occur at
photon energy of $\Delta /2 - (m_i + m_j)^2/(2\Delta)$.
The threshold locations thus determine neutrino pair masses
$m_i + m_j$.
It turns out that the threshold rise of single photon spectrum
is sensitive to neutrino mixing angles such as the unknown
$\theta_{13}$ (see below for more details).
How the momentum conservation holds for
macro-coherent phenomena \cite{rnpe-macro} shall be explained later.

Atoms we consider belong to a class of
$\Lambda-$type three level system,
$|i \rangle_A \,, |n \rangle_A  \,, | f\rangle_A $ with 
energy relation $E_n > E_i > F_f$.
RNPE occurs in second order perturbation,
a combination of weak and QED process, and
has the transition operator of the structure
\cite{my-06},
\(\:
G_F \vec{d}\cdot\vec{E} \vec{S}_e\cdot\vec{j}_{\nu}/(\Delta_{nf}-\omega)
\,,
\:\)
where the electron spin operator
$ \vec{S}_e$ arises from the neutrino pair emission vertex 
for atomic transition $|i \rangle_A \rightarrow | n \rangle_A$, 
multiplied by the neutrino current $\vec{j}_{\nu}$, 
a bilinear form of two plane wave neutrinos,
and $\vec{d}\cdot\vec{E}$ for $|n \rangle_A \rightarrow | f \rangle_A$
is the usual QED E1 
(or M1 $\vec{\mu}\cdot\vec{B}$) vertex.
Here $\Delta_{nf} = E_n - E_f$ is the energy difference 
between two atomic levels, and  $\omega$
is the photon energy.
We may classify the weak vertex as Gamow-Teller (GT)
type, since it involves the electron spin operator.
The Fermi type operator appears as a small relativistic
correction, and it usually gives much smaller rates
than GT, GT/Fermi interference being of order
$10^{-4}$ less usually.
The transition operator for $\nu_i \nu_j$ emission
is multiplied by $\xi_{ij} = U_{ei}^*U_{ej} - \delta_{ij}/2$, with $U_{ei}$
the neutrino mixing matrix \cite{n-mass-matrix}.
For simplicity we assume that the Majorana phases are absent,
thus
\(\:
U_{e1} = c_{12}c_{13}\,,
\;
U_{e2} = s_{12}c_{13}\,,
\;
U_{e3} =  s_{13}\,,
\:\)
with $c_{ij} = \cos \theta_{ij}\,, s_{ij} = \sin \theta_{ij}$
(\cite{majorana phases} for an effect of Majorana phases).

Two energy differences of atomic levels, 
$\Delta_{if} = E_i - E_f \equiv \Delta$ and 
$\Delta_{ni}=E_n - E_i$,
are involved in the transition matrix element
of RNPE; the larger $\Delta$ and the smaller $\Delta_{ni}$, 
the larger the rate is.
Low lying excited levels of rare gas atoms are 
among the best candidate targets.
For instance, an example of
Xe  gives a good candidate of states; 
$| i \rangle_A = 5p^5(^2P_{3/2})6s ^2[3/2]_2$ 
of total electron angular momentum $J = 2$, 
metastable with lifetime O[40] sec,
$|f \rangle_A = 5p^6\, ^1S_0$ the ground state of Xe atom, and
$|n \rangle_A = 5p^5(^2P_{3/2})6s ^2[3/2]_1 $ of $J=1$. 
These low lying excited states are well described by
two "particle" picture of electron and hole; $(6s)(5p)^+$.
The system is nearly degenerate, $\Delta_{ni} \ll 
\Delta $, giving a large RNPE rate.
In isolated environment $\Delta_{ni} \sim 0.12$ eV and
$\Delta \sim 8.4 $ eV for Xe.

Preparation of atomic polarization in excited states is not difficult in
optical experiments using atoms,  made possible by
tuning excitation laser frequency,
if a magnetic field (its direction
taken to $\parallel$ z-axis) is applied to distinguish
Zeeman sublevels.
Both for Xe atoms of even (having no hyperfine
split levels) and odd nuclei (both HFS and Zeeman splitting present)
we may use the total angular momentum
and its component along the magnetic field
$F, M$ to distinguish these energy levels.

A complication that arises to obtain large PV effects
is that one needs a nearly equal mixture of states of magnetic
quantum numbers $M$, different by $\pm 1$.
This is made evident below
in explicit computations. 
There may be a variety of methods to realize this mixture,
and we consider here a simultaneous application of
tilted electric field.
As is shown in Figure 1
for odd $^{131}$Xe of nuclear spin $I = 3/2$ \cite{even xe}, 
the crossing of different hyperfine levels occurs at $\sim$ 1.5 kG for
$| 7/2, -1/2 \rangle $ and $|5/2, 1/2 \rangle$, creating degeneracy
(we consulted \cite{hf crossing} for this computation).
Zeeman splitting due to the magnetic field
is the first order effect of the field,
while the Stark effect due to the electric field
is of the second order in field.
Presence of the weak electric field 
removes the  degeneracy at the level crossing point,
generating a state of nearly equal mixture of states of the form,
\(\:
\cos \eta |F, M \rangle + \sin \eta |F-1, M+1 \rangle
\:\)
of $\tan \eta \sim \pi/4$.
There are actually two nearly degenerate states, this one
and the other orthogonal one.
The orthogonal one gives PV effects exactly opposite in sign,
hence one needs to separate two contributions.
This is experimentally possible,
for instance, by adiabatically changing the magnetic
field strength and choosing either of the two
branches of level repulsion.
Parity violating effects then appear as $\propto \cos \eta \sin \eta$.

\begin{figure*}[htbp]
 \begin{center}
 \epsfxsize=0.5\textwidth
 \centerline{\epsfbox{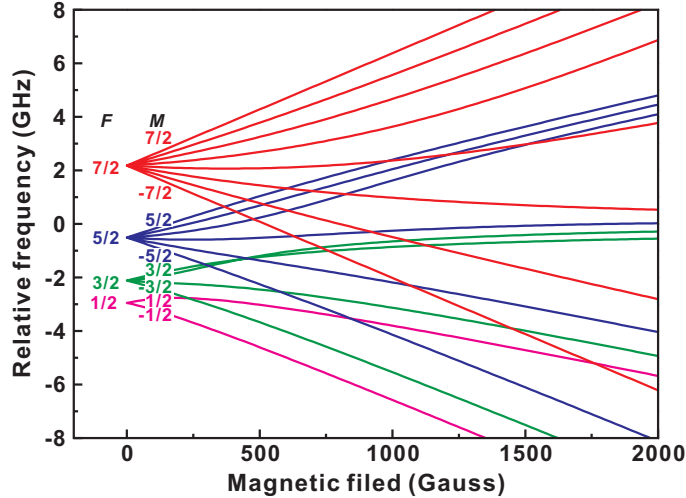}} \hspace*{\fill}
   \caption{Zeeman levels of odd $^{131}$Xe atom.
      }
   \label{fig1:}
 \end{center} 
\end{figure*}

Relevant electronic transitions are of the type
$J=2 \rightarrow J = 1 \rightarrow J =0$ (note $\Delta F = \Delta J$).
Suppose that the magnetic quantum number along z-axis is denoted by $M$.
We wish to obtain mixture of states, $M = -1/2$ and $1/2$ states
for finite PV effects.
The electron hole in the inner core $(5p)^+$
of the initial state has J = 3/2 for
relevant J = 2 $\rightarrow $ 1 transition.
For initial $M = \pm 1/2$ mixture
one may consider a hole in $p_{-}$ and $p_z$
whose wave functions are $\propto (x - iy)/\sqrt{2}$
and $\propto z$.
Suppose that a weak Stark field $E_s$ is applied to a tilted direction
in $(yz)$ plane, with a tilted angle $\psi$ from z-axis.
Second order Stark effects caused by Hamiltonian
$E_s (z \cos \psi + y \sin \psi)$ then leads to
an off-diagonal mixing of the type, 
\(\:
6s (5p_{-})^+ \rightarrow (6s)^2 \rightarrow 6s (5p_z)^+
\:\)
giving a mixture of 
states, $6s (5p_{\pm})^+$ and $6s (5p_z)^+$.
The matrix element for this is
$\propto \cos \psi \sin \psi$, thus 
is maximal at $\psi = \pi/4$.
In the rest of numerical estimates we assume $\psi = \pi/4$.

{\bf  RNPE with invisible neutrino pair}
\hspace{0.5cm}
For illustration 
we consider $^{131}$Xe atomic transitions of the type,
\begin{eqnarray}
&&
\hspace*{-2cm}
\left\{
\begin{array}{l}
\cos \eta | 7/2, -1/2 \rangle  \\
\\
\\
+ \sin \eta | 5/2, 1/2 \rangle \\
\end{array}
\right\}
\rightarrow 
\left\{
\begin{array}{l}
| 5/2, -3/2 \rangle  \\
| 5/2, -1/2 \rangle \\
| 5/2, 1/2 \rangle \\
| 5/2, 3/2 \rangle  
\end{array}
\right\}
\rightarrow
\left\{
\begin{array}{l}
| 3/2, -3/2 \rangle \\
| 3/2, -1/2 \rangle \\
| 3/2, 1/2 \rangle \\
| 3/2, 3/2 \rangle
\end{array}
\right\}
\,.
\end{eqnarray}
Summation over 4 final states is inevitable
in a simplest experimental setup that
does not measure final atomic states.

Neutrinos are difficult to detect, but
one may measure energy, emission angle, and circular polarization 
of the accompanying  photon in RNPE, and derive
event rates of specified photon energy-momentum, $\vec{k}$, and helicity
$h_{\gamma}$,
in the form of continuous spectrum.
On the other hand, we
sum over non-observable variables; neutrino helicities $h_i$ and
their momenta $\vec{p}_i$ of the pair.
The neutrino helicity summation in rates \cite{my-06}
gives  bilnear terms of neutrino momenta; $K^S_{ij} = $
\begin{eqnarray}
&&
\hspace*{-2cm}
\sum_{h_1,h_2}  j_M^{i} (j_M^{j})^{\dagger} 
= 
(- \frac{m_1 m_2}{E_1 E_2}\delta_M + 1)\delta_{ij}
+ \frac{1}{E_1 E_2}
(p_1^i p_2^j + p_1^j p_2^i - \delta_{ij}\vec{p}_1\cdot\vec{p}_2)
\,.
\end{eqnarray}
The case $\delta_M = 1$ is applied for Majorana neutrinos,
and $\delta_M = 0$ for Dirac neutrinos.
The subsequent neutrino momentum integration
(with $E_i = \sqrt{p_i^2 + m_i^2}$ neutrino energy), with
\(\:
\int dP_{\nu} = \int d^3p_1 d^3p_2 \delta^3
(\vec{k} + \vec{p}_1 + \vec{p}_2) \delta (\Delta - \omega - E_1 - E_2)
K^S_{ij}
\,,
\:\)
gives
a 1-dim. neutrino energy integral $\int dE_1$, which
can be done analytically.

The result of neutrino helicity and momentum summation
gives quadratic forms of photon momentum,
$k_i k_j$ and $ \delta_{ij}\vec{k}^2 \times$ functions
of photon energy $|\vec{k}| = \omega$.
These quadratic forms are to be multiplied by
a product of
electron spin operators $S_i S_j$.
Since an approximate azimuthal symmetry around
the magnetic (or atomic polarization) axis exists,
one may integrate, or average, over the azimuthal angle $\varphi$
of emitted photons, $\int_0^{2\pi} d\varphi$.
This allows omitting various bilinears like
$k_z k_x \propto \cos \varphi, k_z k_y \propto \sin \varphi, 
k_x k_y \propto  \cos \varphi \sin \varphi$ which all vanish after
the azimutal angular integration.
Thus, only two types of effective electron spin operators 
\(\:
S_{\pm}S_{\mp} \,, S_z S_z \,,
\:\)
($S_{\pm} = S_x \pm i S_y$) contribute to rate calculation.

For QED vertex we use two independent circular polarizations 
of $h_{\gamma}= \pm 1$,
and write QED vertex amplitude $\vec{d}\cdot\vec{E}$ as circularly
polarized electric field $\times$
\(\:
\frac{1 \pm c}{2} d_+ + \frac{1\mp c}{2} d_- \pm \frac{s}{\sqrt{2}} d_z
\:\)
with $c = \cos \theta, s = \sin \theta$ 
($\theta$ the photon emission angle from polarization axis) and
$d_{\pm} = \mp (d_x \pm i d_y)/\sqrt{2}$.
The squared amplitude $d_i d_j E_i E_j^*$
is then bilinear function of the angle factor $c,s$.

An example of parity violating (PV) term ending at the final state
$|3/2, -1/2 \rangle$ with emission of $h_{\gamma} = \pm 1$ photon gives
a rate proportional to $\cos \eta \sin \eta \,\times$
\begin{eqnarray}
&&
\frac{\pm s(1\pm c)}{2\sqrt{2}} 
\langle 5/2,- 3/2 |j_+ S_- |7/2, - 1/2 \rangle 
\langle 5/2, 1/2 |j_- S_+ |5/2, -1/2 \rangle
\nonumber \\ &&
\times
\langle 3/2, -1/2 | E_- d_+ | 5/2,- 3/2\rangle
\langle 5/2, -1/2 | E_z^* d_z |3/2, -1/2 \rangle 
\,.
\nonumber
\end{eqnarray}
This term gives the angular asymmetry
$\propto \sin \theta \cos \theta$ 
for unpolarized (averaged over helicity $h_{\gamma} = \pm 1$) 
photon emission, asymmetric with respect to
$\theta = \pi/2$.
At the same time the same term gives circular polarization, 
when summed over the helicity weight 
$\sum_{h_{\gamma}} h_{\gamma} \times$ rate,
leading to angular symmetric terms $\propto \sin \theta $.

Thus, unless other added terms cancel in total, two
PV effects remain as $\propto \cos \eta \sin \eta$.

Atomic wave functions we need for this Xe case
are E1 dipole between the intermediate J = 1, and
the final J=0 states, for which we use the
measured decay rate $\gamma_{nf}$ given by
\(\:
\omega^2 \sum_{{\rm pol}}|\vec{d}\cdot{E}|^2/\pi = \omega^3\gamma_{nf}/
\Delta_{nf}^3
\:\).
The initial and the intermediate wave function
overlap is due to the electron spin flip
caused by the neutrino pair emission, hence having
a large electronic orbital overlap.
This part can be worked out approximately by using Clebsch-Gordon (CG) coefficients
\cite{rose} alone.

\vspace{0.5cm}
{\bf Spectrum rates and PV observables}
\hspace{0.5cm}
A general rate formula may be
given under specified photon circular polarization $h_{\gamma}= \pm 1$,
\begin{eqnarray}
&&
\hspace*{-1cm}
\frac{d^2\Gamma }{d\omega d\cos \theta }(\omega, h_{\gamma})
= \frac{G_F^2 \gamma_{nf} nN \omega^3 }
{ 8\pi^2 \Delta_{nf}^3(\Delta_{nf} - \omega)^2 }  \sum_{ij} |c_{ij}|^2
\left( b_4 c^4 + b_2 c^2 + b_0 + s (a_3 c^3 + a_1 c)
\right.
\nonumber \\ &&
\left.
+ h_{\gamma}( e_3 c^3 + e_1 c + s(f_2 c^2 + f_0 )\,) \right) \,,
\label{general rate fomula 1}
\end{eqnarray}
where $c = \cos \theta, s = \sin \theta$.
Here  $a_i, b_i, e_i, f_i$ of mass dimension 2 are
functions of variables $\omega$
and elements of the neutrino mass matrix,
containing product of 4 CG coefficients.
Parity violating effects appear in
$a_i$ and $f_i$ terms $\propto \cos \eta \sin \eta$.
Parity conserving terms $\propto b_i, e_i$ are multiplied by
$\cos^2 \eta $ or $\sin^2 \eta$.

Without measurement of circular polarization
$h_{\gamma}$,
there are components, symmetric one (even function of $\cos \theta$)
relative to the
atomic polarization axis,
and asymmetric one (odd function of $\cos \theta$).
The total rate $\propto b_4/5 + b_2/3 + b_0$
is an integral of symmetric functions.
The angular asymmetry gives parity violating effect;
for instance, the  difference
in the forward (F) and the backward (B) hemispheres
at some particular angle $\theta$ and $\pi - \theta$
of $0 \leq \theta \leq \pi/2$.
We plot in Figure 2 the angular
asymmetry defined by the ratio of FB difference
to FB sum,
\(\:
s(a_3 c^3  + a_1 c )/(b_4 c^4 + b_2 c^2 + b_0) \,,
\:\),
taking several $\theta$ of $\pi/10, \pi/6, \pi/4, \pi/3, 2\pi/5$
(with $ \eta=\pi/4$).

\begin{figure*}[htbp]
 \begin{center}
 \epsfxsize=0.5\textwidth
 \centerline{\epsfbox{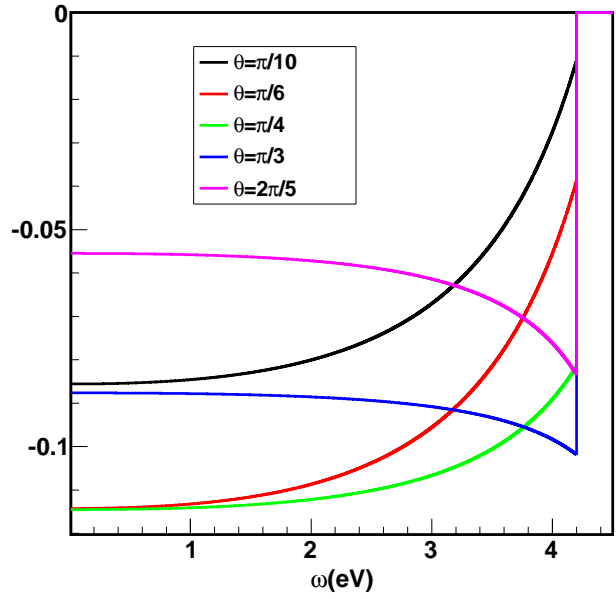}} \hspace*{\fill}
   \caption{Angular asymmetry of emitted photon at several angles $\theta$
   from  oriented $^{131}$Xe atoms in a metastable $J=2, F=7/2$ state
   as a function of
   photon energy $\omega$.
   Assumed neutrino masses are 1meV, 10 meV, 50meV,
  with mixing angles, $\sin^2 \theta_{12} = 0.3 \,, 
   \sin^2 \theta_{13} = 0.039$ assuming the normal
   hierarchy \cite{n-mass-matrix}.
   }
   \label{fig2:}
 \end{center} 
\end{figure*}

With measurement of circular polarization,
the simplest P-odd quantity is the relative circular polarization as 
a function of angle summed at two
angles $\theta$ and $\pi - \theta$, given by
\(\:
s(f_2 c^2 + f_0 )/(b_4 c^4 + b_2 c^2 + b_0) \,.
\:\)
We plot in Figure 3 their spectrum  (taking $ \eta=\pi/4$)
at the same angles as in Figure 2.

\begin{figure*}[htbp]
 \begin{center}
 \epsfxsize=0.5\textwidth
 \centerline{\epsfbox{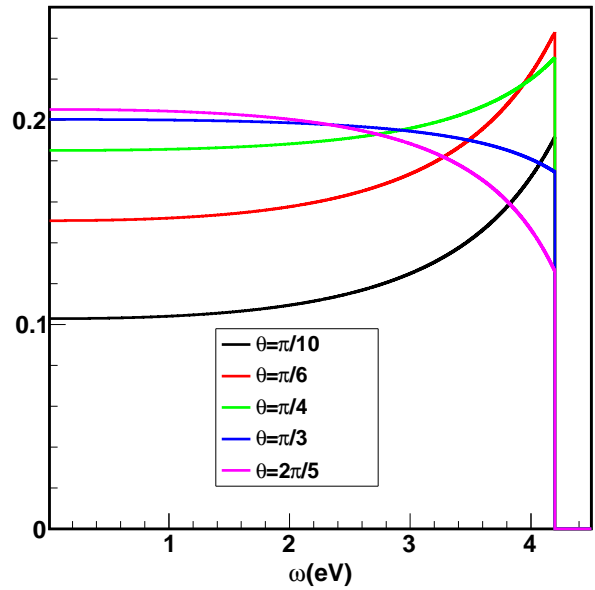}} \hspace*{\fill}
   \caption{Circular polarization for the Xe metastable state
    at several angles
    from oriented atoms.
   Assumed parameters are the same as in Figure 2.
   }
   \label{fig3:}
 \end{center} 
\end{figure*}

Both of these PV quantities are of order unity,
typically $O[0.1]$,
even far away from neutrino mass thresholds,
since parity violation is intrinsic to
the involved weak interaction not related 
to the finiteness of neutrino masses.
The feature of large parity violation
far from thresholds where the event rate is larger 
can be used for rejection of QED 
backgrounds and a positive identification of RNPE.
The magnitude of PV quantities differ significantly
at different emission angles.

Parity-odd quantity thus computed is almost of equal magnitude in the Majorana
and the Dirac cases, the difference being $\leq O[2\times 10^{-4}]$ in
Xe case of a large energy difference $\sim$ 8.4 eV.
The Majorana vs Dirac distinction using P-even
quantity is made easier by using atomic
system of smaller energy difference close to
neutrino pair masses such as $2m_3$ (the largest pair mass);
the rate difference may become several $\times 10^{-2}$
for $\Delta <$ 1eV.

\vspace{0.5cm}
{\bf Macro-coherent RNPE}
\hspace{0.5cm}
Coherence of targets is important, both to enhance the event rate
and to determine the threshold location.
Ordinarily, the momentum conservation in elementary
processes 
does not play important roles in atomic
radiative decay, since atomic recoil is very small
and difficult to measure.
We however utilize the macro-coherence of target atoms,
similar to, but more enhanced in rates than
(hence called the macro-coherence), 
a single photon superradiance \cite{sr}.

In macro-coherent phenomena more than two light particles
such as $\gamma \gamma$ in two photon emission
and  $\gamma \nu_i \nu_j$ 
in RNPE are involved
in the final state of decay product.
If the phases of emitted plane-wave
particles of product amplitude $e^{i\sum_i \vec{k}_i\cdot\vec{x}_d}$ 
from various target sites at $\vec{x}_d$ 
(distributed uniformly with a number density $n$) are coherently added,
one has the macro-coherent rate in the large target volume ($V$) limit
\cite{discrete limit},
\(\:
\Gamma = n N
\int \left(
\Pi_i \frac{d^3 k_i}{(2\pi)^3}\right)
(2\pi)^4 \delta(\Delta - \sum_i \omega_i)\,\delta(\sum_i \vec{k}_i)
\,|{\cal M}( \vec{k}_i)|^2
\,,
\:\)
($N = nV$), a result of coherence not restricted by 
the wavelength of emitted photon, quite unlike 
in the single photon superradiance \cite{sr}.
Thus, the coherence prevails in a macroscopic target region,
and 
one expects the momentum conservation,
$\vec{k} + \vec{p}_1 + \vec{p}_2 = 0$ for RNPE,
and a large enhanced rate.

\begin{figure*}[htbp]
 \begin{center}
 \epsfxsize=0.5\textwidth
 \centerline{\epsfbox{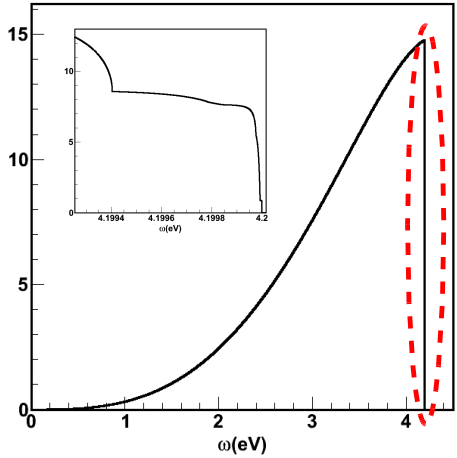}} \hspace*{\fill}
   \caption{RNPE spectrum rate for the $^{131}$Xe metastable state
    in the unit of $nN G_F^2 \gamma_{nf}/(2\pi^2 \Delta_{nf}^3)$ of eq.(\ref{general rate fomula 1}).
   The red-dotted
   threshold region including all six is expanded in the inset
   to exhibit threshold rises.
      Assumed parameters are the same as in Figure 2.
     }
   \label{fig4:}
 \end{center} 
\end{figure*}

Coherence dynamically develops, and the key for this is 
the seed of quantum fluctuation
of fields  or succesive spontaneous emission
within the active target volume, as emphasized
in superradiance phenomena \cite{sr}.
Moreover, an active external trigger by a pulsed laser
may  help much for the coherence development \cite{triggered sr}.
Analysis based on the Maxwell-Bloch equation \cite{sr}
is expected to give strong angular correlation along the axis
of exciting and triggering laser, 
an effect not evident in the present argument
and ones in \cite{rnpe-macro} as well.

The total rate $\int d\omega d\Gamma/d\omega$
from the entire target roughly scales with
\(\:
nN G_F^2 \gamma_{nf}/(2\pi^2 \Delta_{nf}^3 ) 
\:\)
times an integrated number.
This is numerically
\( \:
\sim 4 \times 10^{-9} (n/10^{24}{\rm cm}^{-3}) (N/10^{20})
(\gamma_{nf}/10^9 s^{-1})
\:\)
Hz times an integrated spectrum in the eV$^4$ unit.
A 100 gr of solid $^{131}$Xe gives the total rate 
of order $\sim 3 \times 10^{-5}$ Hz,
taking into account the enhancement due to
the soliton formation crudely estimated in \cite{ys-09} to be $O[10^4]$.
One needs a more detailed numerical integration
of Maxwell-Boltzmann equation to give a more accurately
determined rate.

The elementary rate from a single atom
may be estimated by taking $N=1$, and replacing the target number density $n$
by the phase space density $\sim \Delta^3$
of atomic decay, which is $\sim$wavelength$^{-3}$.
Thus, the macro-coherent rate is enhanced by the factor of the ratio
of target volume/wavelength$^{3}$ compared to the elementary rate.
In contrast the single photon superradiance has the enhancement factor
of order, (linear dimension/wavelength)$^2$ \cite{sr}.

We show in Figure 4 the total spectrum rate, taking $ \eta=\pi/4$.
A global feature of the macro-coherent spectrum is
the sharp rise from thresholds and a fast decrease towards
small photon energy region, ultimately $\propto \omega^3$.
The step around 4.1994 eV is due to (33) mass threshold, whose 
location, but not spectral shape, is insensitive to the angle factor 
$\theta_{13}$ if $\sin^2 \theta_{13} \leq 0.039$ (the present limit).
In order to locate thresholds one needs accuracy of
photon energy resolution of order $10^{-3}$ or even less.
This accuracy is not problematic, since one can trigger RNPE
by laser whose energy resolution is far better.
The fraction that threshold regions occupy is however not large;
for instance the ratio of integrated rate of the inset part
to the total in Figure 4 is 
$\sim 5 \times 10^{-4}$ for all 6 threshold regions.
We also worked out the case of inverted hierarchy of neutrino masses;
the difference between the two cases is not overwhelming,
but a sizable spectral shape difference exists between
the two cases, the normal and the inverted hierarchies.
In another word one can equally investigate and experimentally distinguish
the normal and the inverted cases by this type of atomic
experiments.

In this Xe example it is necessary to obtain large high statistics
data near threshold regions for the precision neutrino mass spectroscopy.
This is due to that Xe level spacing $\sim 8.4$eV is somewhat large
compared to the anticipated neutrino mass $\sim 0.05$eV
in the assumed normal hierarchy case.
Atoms of much smaller energy spacing are far better
for the purpose of precision neutrino mass spectroscopy.

To obtain a large enhanced macro-coherent rate,
one needs a collection of atomic targets of O[Avogadro number]
${\rm cm}^{-3}$,
well isolated
among themselves and weakly interacting with envirornment.
An example is a matrix implantation of a fraction of targets
within solid matrix such as para-H$_2$.

\vspace{0.5cm}
In summary,
we discussed two types of parity violating effects;
angular asymmetry from oriented atoms and
emergent circular polarization,
as a means to positively verify weak RNPE process, which can
be used to systematically determine the neutrino mass matrix.

\vspace{0.5cm}
{\bf Acknowledgements}
\hspace{0.5cm}
We should like to thank our collaborators
of SPAN group, K. Nakajima, I. Nakano,
and H. Nanjo for enlightening discussions on 
experimental aspects of related subjects.


\begin{thebibliography}{99}
\bibitem{majorana phases}
Effects of the Majorana phases $\alpha\,, \beta$ are
to change two $U_{ei}$'s to $U_{e2} = s_{12}c_{13}e^{i\alpha}\,, 
U_{e3} = s_{13}e^{i\beta}\,.$
The amplitude of neutrino pair emission contains
both imaginary and real parts of $\xi_{ij}^2$ 
($\xi_{ij}$ defined in the text).
Time-reversal even terms which we are concerned with
depend on $|\xi_{ij}|^2$ (contributing to 
commeon terms in Dirac and Majorana cases)
and $(\Re \xi_{ij})^2 - (\Im \xi_{ij})^2$ (Majorana interference
term).
The Majorana phases $\alpha, \beta$ thus appear in
the Majorana interference rates at (12),
(13), (23) thresholds, in the form of
$c_{12}^2s_{12}^2c_{13}^4\cos 2 \alpha\,, 
c_{12}^2c_{13}^2s_{13}^2\cos 2 \beta\,, 
s_{12}^2c_{13}^2s_{13}^2\cos 2 (\alpha - \beta)$, respectively.
Their non-vanishing effects are relatively small
except at (12) threshold in which no small factor $\sin^2 \theta_{13}$ exists.
\bibitem{even xe}
The problem with Xe of even nuclei is that the
state mixture occurs for two states separated by
different $J$ with a large energy difference, hence 
the mixing angle is expected very small.
\bibitem{discrete limit}
However, if phases cancel by some other reasons,
as in the dilute gas,
the incoherent sum results, giving the rate $ N $ 
(the total target number) $\times$  the elementary 
rate for a single atom without the momentum conservation.
Thus, on general grounds,
one cannot take the dilute gas limit in the formula of macro-coherent rate.
\end{thebibliography}
\end{document}